\documentclass[twocolumn]{aa}  
\usepackage{graphicx}
\usepackage{natbib}
\usepackage{color}           
\definecolor{mgc}{RGB}{0,0,0}
\newcommand{\mgc}[1]{{\color{mgc} #1}}
\definecolor{mrk}{RGB}{0,0,0}
\newcommand{\mrk}[1]{{\color{mrk} #1}}
\definecolor{dltd}{RGB}{0,0,0}

\begin{document}

\title{Polarisation of microwave emission from reconnecting twisted coronal loops}

   \author{M. Gordovskyy \inst{1}\fnmsep\thanks{\email{mykola.gordovskyy@manchester.ac.uk}}, 
	  P.K. Browning\inst{1}   
          \and
	E.P. Kontar\inst{2}}

   \institute{Jodrell Bank Centre for Astrophysics, University of Manchester, Manchester M13 9PL, UK\\
         \and
             Astronomy \& Astrophysics Group, University of Glasgow, Glasgow G15 8QQ, UK\\
             }

   \date{Received ; accepted }

\authorrunning{Gordovskyy et al.}
\titlerunning{Microwave emission from twisted loops}

 
  \abstract
   {Magnetic reconnection and particle acceleration due to the kink instability in twisted coronal loops can be a viable scenario for confined solar flares. Detailed investigation of this phenomenon requires reliable methods for observational detection of magnetic twist in solar flares, which may not be possible solely through extreme UV and soft X-ray thermal emission. Polarisation of microwave emission in flaring loops can be used as one of the detection criteria.}
   {The aim of this study is to investigate the effect of magnetic twist in flaring coronal loops on the polarisation of gyro-synchrotron microwave (GSMW) emission, and determine whether it could provide a means for magnetic twist detection.}
   {We use time-dependent magnetohydrodynamic and test-particle models developed using LARE3D and GCA codes to investigate twisted coronal loops relaxing following the kink-instability. Synthetic GSMW emission maps (I and V Stokes components) are calculated using GX simulator.}
   {It is found that flaring twisted coronal loops produce GSMW radiation with a gradient of circular polarisation across the loop. However, these patterns may be visible only for a relatively short period of time due to fast magnetic reconfiguration after the instability. Their visibility also depends on the orientation and position of the loop on solar disk. Typically, it would be difficult to see these characteristic polarisation pattern in a twisted loop seen from the top (close to the centre of the solar disk), but easier in a twisted loop seen from the side (i.e. observed very close to the limb).}
   {}


   \maketitle
%

\section{Introduction}\label{intro}

\begin{figure}[ht!]
\centerline{\includegraphics[width=0.34\textwidth,clip=]{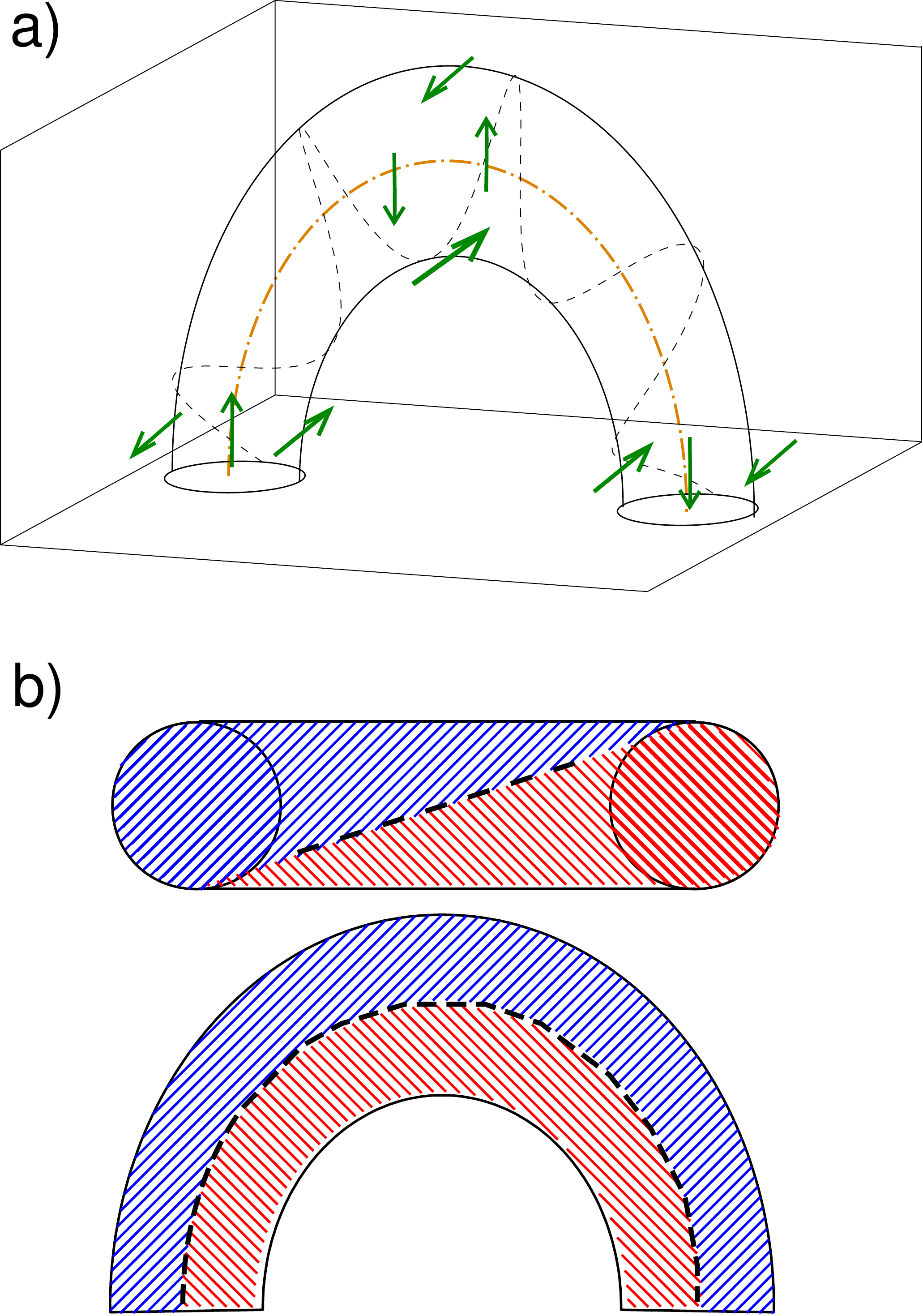}}
\caption{Panel (a): schematic drawing of a twisted loop. Green arrows show the longitudinal and azimuthal magnetic field in selected locations. Orange dot-dashed line denotes the "skeleton line", the magnetic field line connecting centres of loop footpoints. Panel (b) shows schematic drawings of the cross-loop polarisation gradient (CLPG) patterns in cases when a loop is observed from its top and from its side. Blue $/$ hatching and red $\backslash$ hatching correspond to positive and negative values of Stokes V, respectively. Black dashed lines denote Stokes V$=0$.}
\label{f-sketch}
\end{figure}

\begin{figure}[ht!]   
\centerline{\includegraphics[width=0.42\textwidth,clip=]{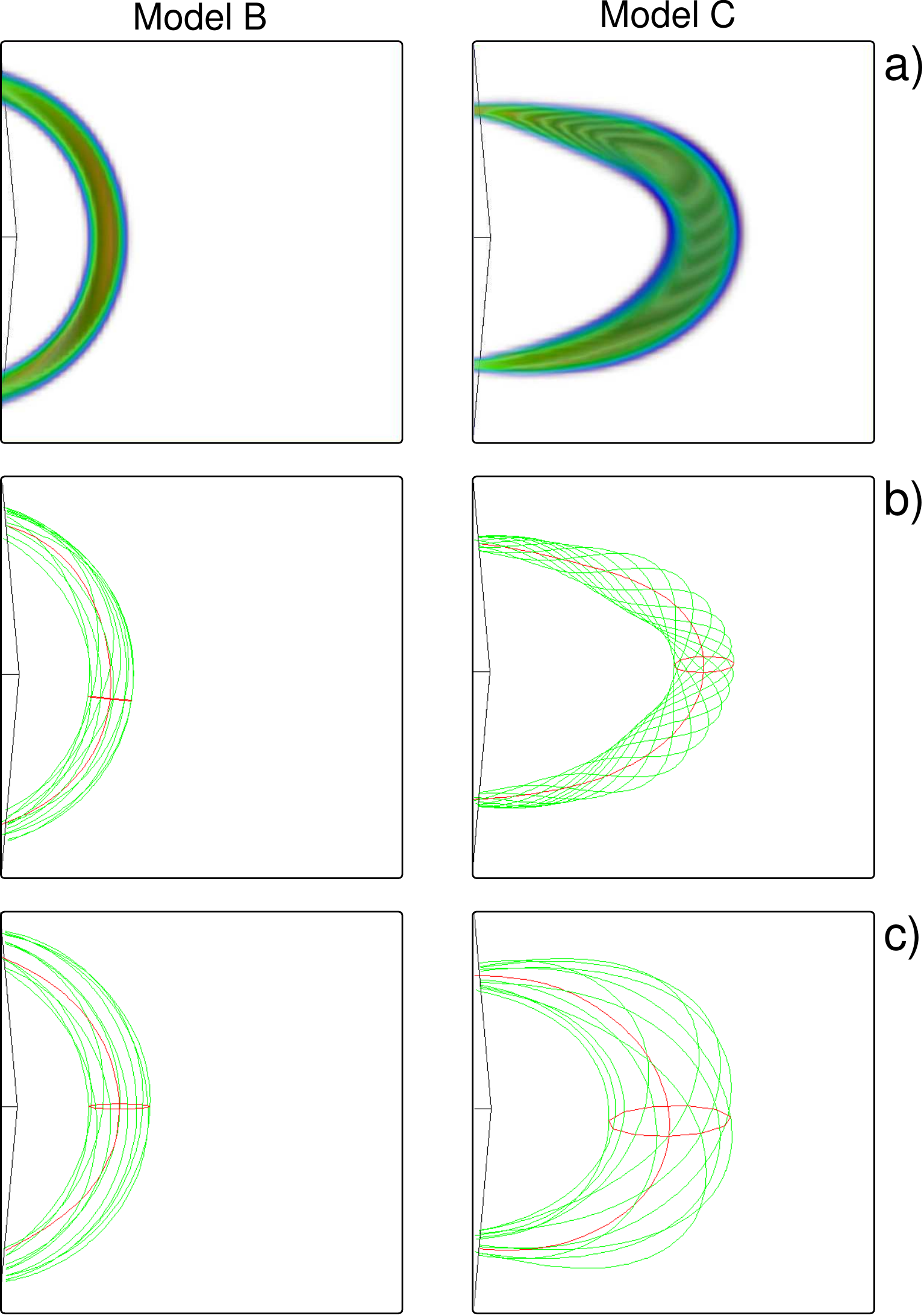}}
\caption{Evolution of loops in models with weak convergence (left column, model B) and strong convergence (right column, model C). Row (a) show spatial distributions of energetic particles in these two models during the onset of magnetic reconnection. Rows (b) and (c) correspond to t=2~s and t=58~s, respectively.}
\label{f-evolv}
\end{figure}

Reconnecting twisted coronal loops are a good alternative to the standard model \cite[e.g.][]{shie95} for explaining some types of solar flares, particularly smaller flares observed in isolated coronal loops \citep{asce09}, and erupting flaring coronal loops \cite[e.g.][]{fan10,kakl10}, including ``failed eruptions'' \citep{alee06,kure13}. One of the key benefits of the energy release scenario involving the kink-instability and magnetic reconnection in twisted loops is energy release and particle acceleration which are distributed within the flaring loop volume \cite[see also discussion in][]{gobr12,gore14}. 

There are numerous observations of twisted loops in solar flares, usually as an element of a major flare \cite[e.g.][]{srie10}. Twisted threads in flaring loops and strong azimuthal magnetic field around loop foot-points are considered to be indicators of magnetic twist. However, a detailed study of this phenomenon requires analysis of twisted loops in solar flares of different sizes and with different configurations, including flares occuring in complex active regions, smaller flares etc. This, in turn, require a reliable methods for observational detections of flaring twisted coronal loops. Recently, possible observational manifestations of magnetic twist in reconnecting coronal loops have been studied using realistic models of energy release in kink-unstable twisted loops in a stratified corona. These models combine magnetohydrodynamic (MHD) and test-particle approaches to represent the thermal and non-thermal components of plasma, respectively  \citep{gobr11, gore14, bare16}. Particularly, these studies address thermal soft X-ray (SXR) and extreme ultra-violet (EUV) continuum emissions, non-thermal hard X-ray emission, and shapes and positions of EUV coronal lines. Firstly, it is shown that some twist should be seen in SXR and EUV thermal emission, although it will be substantially lower than the critical twist, leading to the kink instability \citep{pine16}. Secondly, it is shown that the sizes of HXR sources produced by energetic electrons in the reconnecting twisted loop should increase with time \citep{gobr11,gore14}. This effect is consistent with RHESSI observations \citep{kone11,jeko13}. Finally, it is shown that turbulent broadening and Doppler shifts of EUV spectral lines produced by reconnecting twisted loops correlate with plasma temperature \citep{gore16}. The non-thermal broadening of spectral lines following from these models is consistent with observations \citep{dose07,dose08}; however, this phenomenon is likely to be observed in other magnetic configurations and cannot be used for observational detection of twisted loops. The gradual increase of the HXR foot-points is more specific and, in principle, can be used for observational detection, alongside other effects, provided the foot-points can be resolved, which can be done for large loops with cross-sections of at least few Mm. Expectedly, a twist visible in thermal EUV emission seems to be the most reliable detection feature. However, it would not be necessarily present in all twisted loops: for this effect, a loop has to contain threads with different emissivity (for instance, due to different temperature or plasma density) which is not always the case.  Therefore, additional criteria are necessary for reliable observational detection of these magnetic field configurations.

\begin{figure*}[ht!] 
\centerline{\includegraphics[width=0.6\textwidth,clip=]{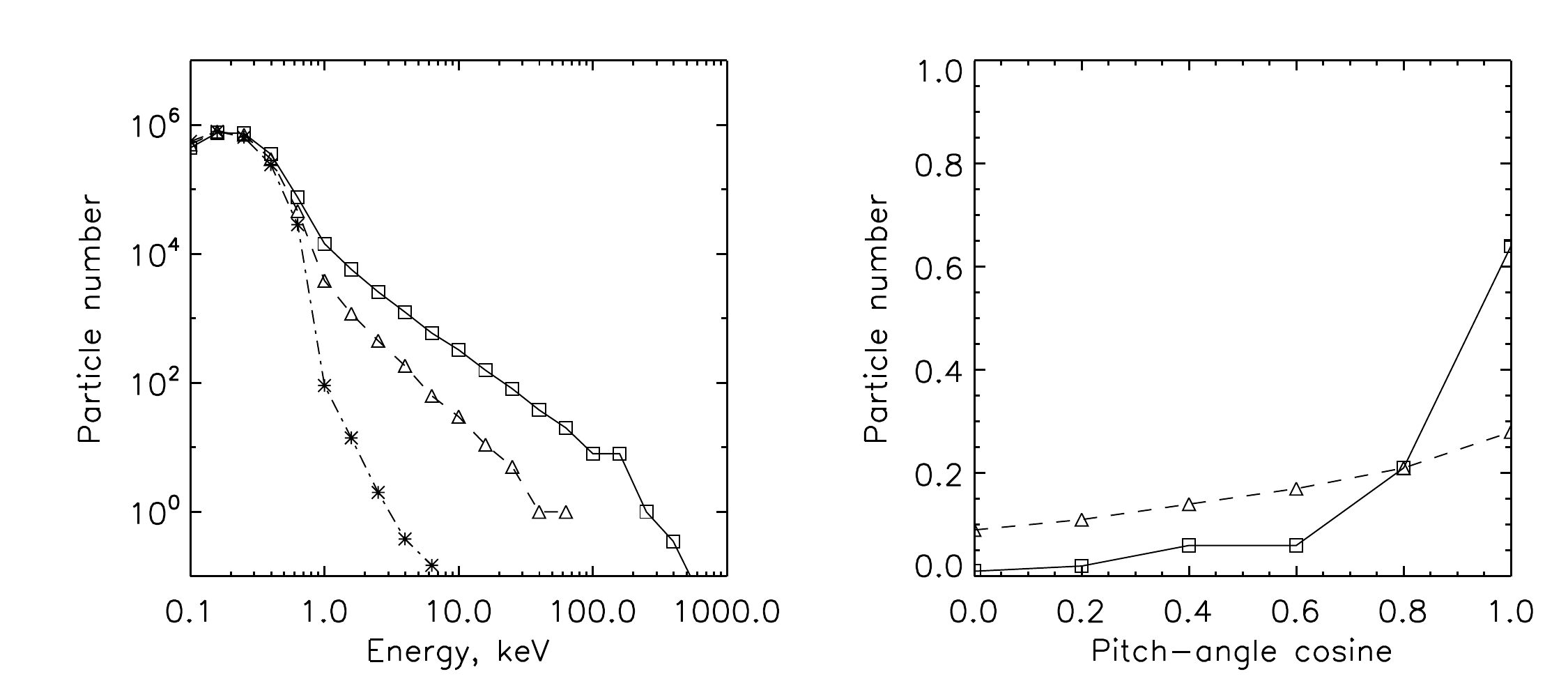}}
\caption{Energy distributions of particles (left panel) and pitch-angle cosine distribution of particles with energies above 10~keV (right panel) in Model B. Solid line with squares corresponds to t=2~s (just after the onset of reconnection), dashed line with triangles corresponds to t=30~s (the moment of highest plasma temperature), and dot-dashed line with stars corresponds to t=58~s (towards the end of reconnection). There is no pitch-angle distribution for t=58~s , because there are not enough particles with energies above 10~keV. Particles numbers are in arbitrary units.}
\label{f-pdistr}
\end{figure*}

Microwave emission is one of the key instruments for solar flare diagnostics. GSMW emission from flares and other active coronal features was extensively studies in the last two decades through forward modelling \cite[e.g.][]{kuce93,nine00,reze15}. Simultaneous fitting of microwave and hard X-ray observations with synthetic spectra appears to be efficient in deriving parameters of non-thermal component of solar flares \cite[for instance][]{tzae08,gime09}.
Gyro-synchrotron microwave (GSMW) radiation produced by electrons in magnetic field is very sensitive to the field direction \cite[e.g.][]{petr81,gare13}. Its circular polarisation (Stokes V component) depends on sign of the line-of-sight magnetic field component. Thus, it is known that opposite foot-points of a flaring loop demonstrate opposite GSMW circular polarisations \cite[e.g.][]{hana05,iwsh13}. Similarly, high-energy electrons in a twisted magnetic loop with strong azimuthal field will produce intence microwave emission in the directions perpendicular to the loop and, most importantly, the sign of circular polarisation (or the sign of V Stokes component) will change across the loop, forming cross-loop polarisation gradient (CLPG) patterns (Figure~\ref{f-sketch}). Therefore, using GSMW polarisation is a natural candidate for a diagnostic of the magnetic field direction and, in particular, for magnetic twist detection.

Recently, \citet{shku16} considered polarisation of stationary and evolving twisted ropes and showed that the CLPG patterns should be clearly visible in most cases. However, a possible difficulty with using this effect for the observational detection of twisted loops might be its life time. Indeed, intense GSMW in flares is generated by high-energy non-thermal electrons and thermal electrons of very hot plasmas. Both appear after the reconnection begins. At the same time, once the magnetic reconnection begins, the twisted loops quickly lose their regular, rope-like shape. Furthermore, in some cases the polarisation degree (the V/I ratio) or the overall intensity of microwave emission may be too low to be observed with available instruments. However, this will depend on a number of factors, such as the life-time of the event (affecting the integration time), complexity of the flaring active region and others.

Therefore, important questions here are (a) how long the CLPG patterns can be observed in flaring twisted loops with typical coronal parameters, and (b) how intense is the microwave emission from these twisted loops and how strong is the polarisation in CLPG pattern. In this paper we investigate microwave emission and its polarisation produced by thermal and non-thermal plasma in evolving reconnecting twisted coronal loop. We calculate GSMW emission from reconnecting twisted coronal loop using our earlier models \cite[e.g.][]{gobr11,bare16}. We use the time-dependent magnetic field and plasma temperature and density from our MHD simulations, along with the energetic electron parameters derived from our test-particle models to calculate microwave emission (I and V Stokes profiles) using the GX simulator developed by \citet{nite15}. The MHD and test-particle models are described in Section~\ref{models}, and the synthetic microwave maps are discussed in Section~\ref{mwave}.

\begin{figure}[ht!]
\centerline{\includegraphics[width=0.28\textwidth,clip=]{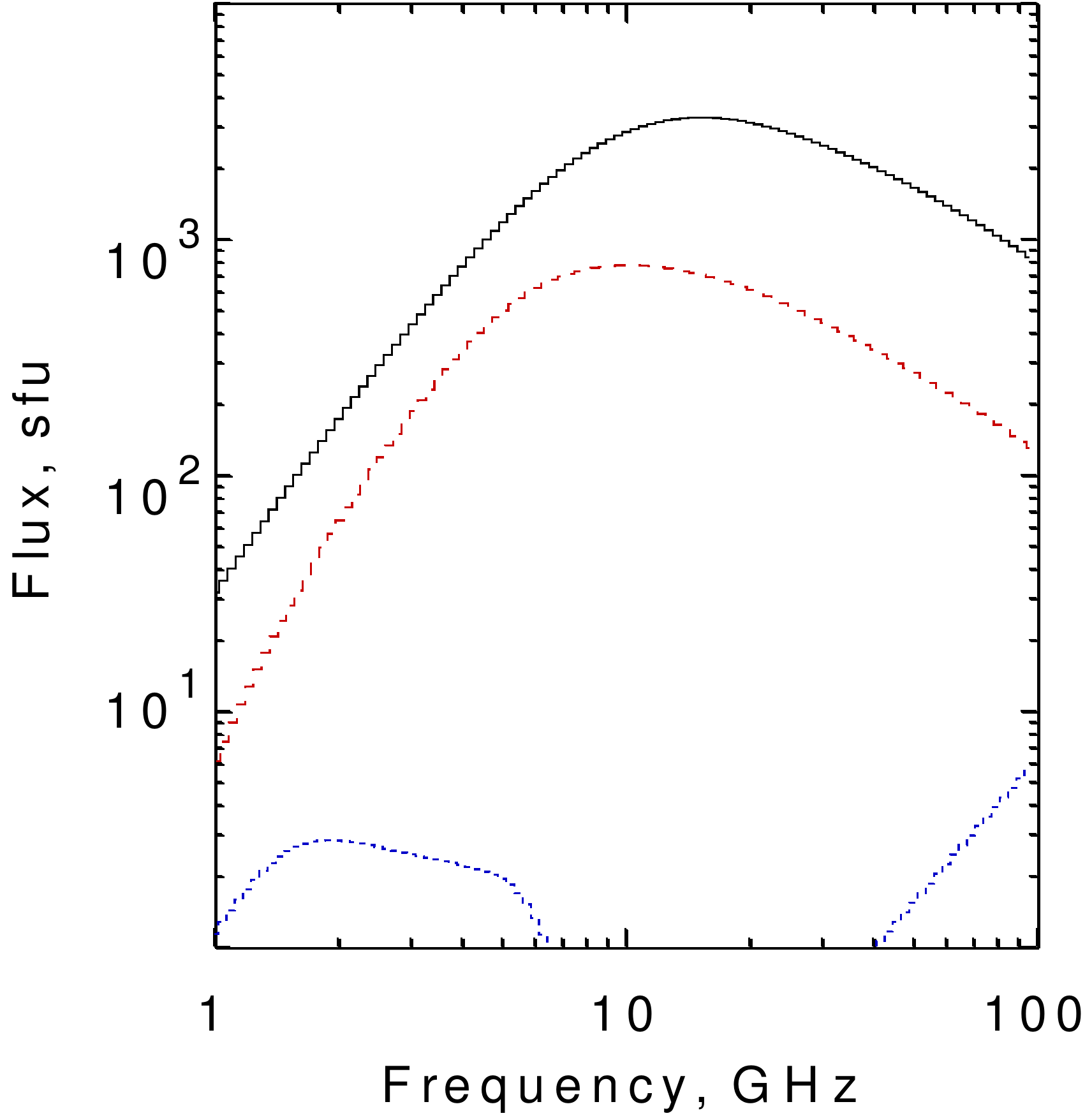}}
\caption{GSMW spectra for the model A for three different moments. Solid black, dashed red and dotted blue lines correspond to t=2~s, t=30~s and t=58~s, respectively.}
\label{f-gsmwspectra}
\end{figure}

\section{Model description}\label{models}

\subsection{Reconnecting twisted loop models}

The evolution of magnetic field and thermal plasma, as well as particle acceleration in reconnecting twisted loops in a stratified atmosphere following kink instability has been described by \citet{gore14,gore16}, \citet{bare16} and \citet{pine16}. Lare3d code \cite{arbe01} has been used for MHD simulations, while GCA code \cite{gore10} has been used for particle simulations.

In the present study, we use three models developed by \citet{gore14,gore16}, with the loop length of about 80~Mm and cross-section radius (near foot-points) of about 4~Mm. We do not consider smaller loops (with the length of $\sim$20~Mm and radius $\sim$1~Mm) because their widths are less likely to be resolved by existing instruments. Due to the properties of the initial potential field, the loops with high magnetic field convergence (foot-point field to loop-top field ratio 10) have relatively higher loop-tops compared to loops with lower field convergence (convergence factor 2) (magnetic field plots are available in Figure~\ref{f-evolv}). Another important difference is that the angle between the magnetic field near footpoints and the boundary (representing the photosphere) depends on the convergence: it is nearly a right-angle for strongly-converging loops, but only about 45 degrees for weakly-converging loops.

The kink instability occurs when the field line twist angle is 4-8$\pi$, depending on the configuration \cite[see][for more detail]{bare16}. What happens after the kink instability has been schematically descibed by \citet{gore14} (see also Figure 6 in that paper). Sudden,  localised increase of the current density, as well as the switching-on of the current-driven anomalous resistivity, result in fast magnetic reconnection and magnetic energy release. The reconnection within the loop results in the reduction of twist, while the reconnection between the twisted loop magnetic field and non-twisted, ambient field results in gradual increase of the loop cross-section. Gradually, the loop loses its rope-like structure and becomes more chaotic. During this time, the current density distribution loses its regular shape becoming very fragmented, rather uniform distribution of small current islands. After this, the loop demonstrates some contraction. 

Spatial distributions and energy spectra of accelerated electrons are calculated using the test-particle simulations with $10^5$--$10^6$ electrons. The initial test-particle population has uniform spatial distribution and isotropic Maxwellian velocity distribution with uniform temperature 0.8~MK. Particles in the flaring loop move predominantly along the field between the two loop footpoints. Most particles accelerated to the energies below $\sim$100~keV thermalise due to Coloumb collisions even before reaching the lower boundary. However, if a particles reaches one of the boundaries, it is allowed to leave the domain; another particle is injected into the domain at the same location, with velocity and pitch-angle taken randomly from the isotropic Maxwellian distribution. (This is known as a ``thermal bath'' boundary condition.)

Due to the magnetic field convergence near foot-points, some particles with low pitch-angles bounce between the opposite foot-points. Some of the electrons experience non-zero electric field in the current sheets associated with magnetic reconnection and get accelerated or decelerated. The typical time of thermal electron passage through the whole domain (or along the whole loop) is 1-10~s, or about one order-of-magnitude shorter than the energy release time; for accelerated electrons, this time is only about 0.1-1~s. 

Only a small fraction of electrons become non-thermal, about 2-4\% of the total particle number during the early stages of magnetic reconnection, then steadily decreasing with time. The energy spectra of accelerated electrons are hard: the electron energy distribution is nearly a power-law between 10-100~keV with the spectral index of $\sim$1.8--2.2 at the onset of energy release and 3.0--3.5 towards the end of reconnection. The accelerated particles are collimated along the magnetic field, {\it i.e.} have pitch-angle cosines distributed around to $\pm 1$ \citep{gobr11}. The pitch-angle distributions are narrower for higher energies. This is, obviously, a consequence of the parallel electric field acceleration: this strongly increases the value of the parallel velocity, while the parpendicular velocities remain nearly thermal.  However, the presence of collisions, and magnetic mirroring (and, hence, a loss cone) in the lower atmosphere makes the distribution more isotropic \citep{gore14}. Because the accelerated electrons move very quickly along the loop, it is difficult to see any spatial structure in particle distribution along the loop \cite[see Figure 11 in][]{pine16}, apart from some increase in non-thermal electron density close to foot-points because of the lower $v_{||}$ velocities due to the magnetic mirroring. 

\begin{figure*}[ht!]    
\centerline{\includegraphics[width=0.7\textwidth,clip=]{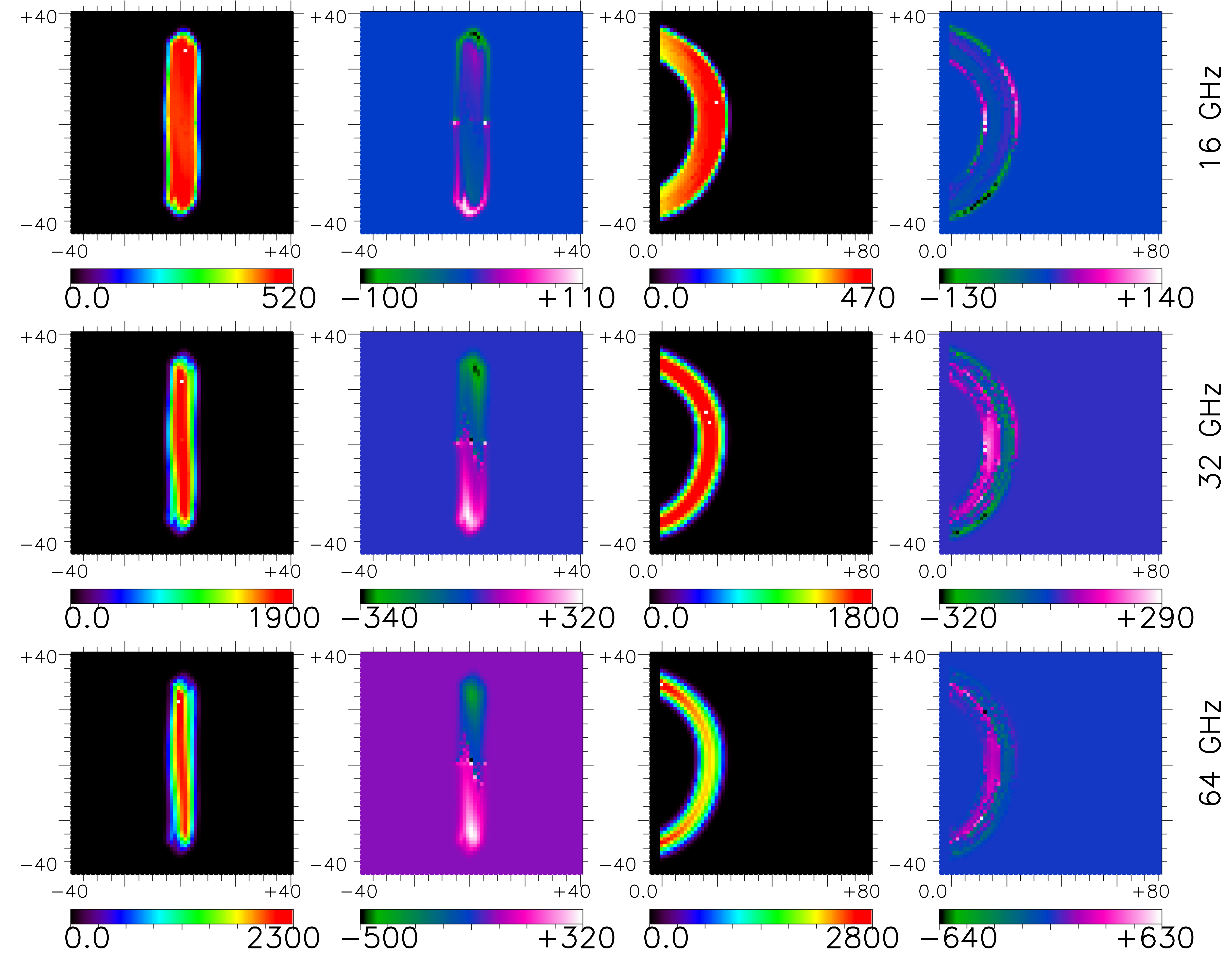}}
\caption{Microwave emission maps from the loop in model B soon after the kink instability and start of the magnetic energy release and particle acceleration (t=2~s). First and second columns correspond to Stokes I and V intensities, respectively, when the loop is seen from the top (X-Y plain). Third and fourth columns are Stokes I and V intensities, respecively, for the loop seen from its side (Z-Y plain). Different rows correspond to different frequencies. Intensities are given in sfu per pixel units, lengths are given in Mm.}
\label{f-so}
\end{figure*}
\begin{figure*}[ht!]    
\centerline{\includegraphics[width=0.7\textwidth,clip=]{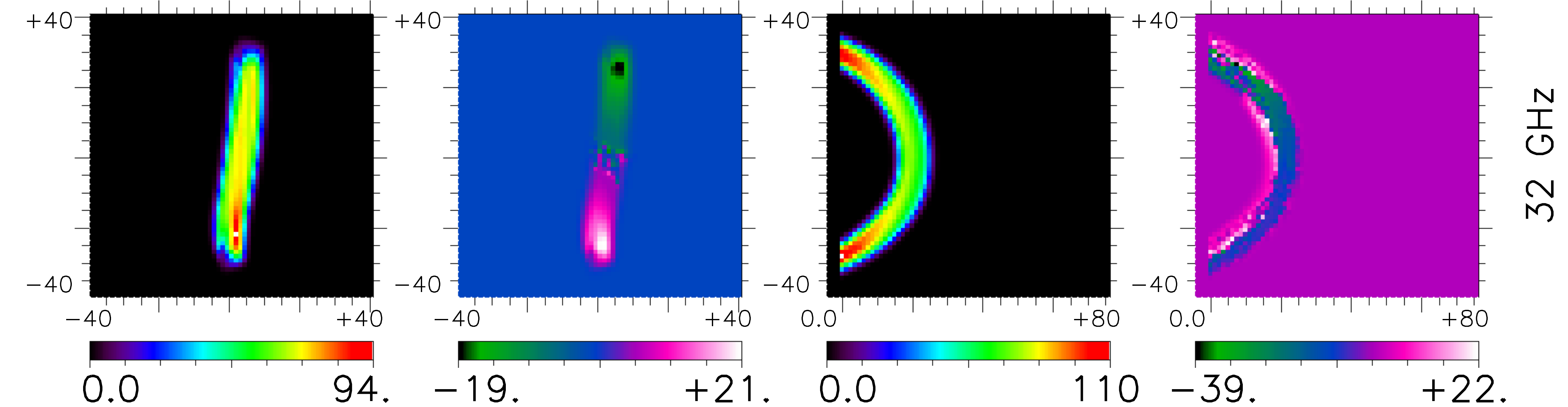}}
\caption{The same as in Figure~\ref{f-so}, but for the moment of peak temperature (t=30~s).}
\label{f-sf}
\end{figure*}
\begin{figure*}[ht!]   
\centerline{\includegraphics[width=0.7\textwidth,clip=]{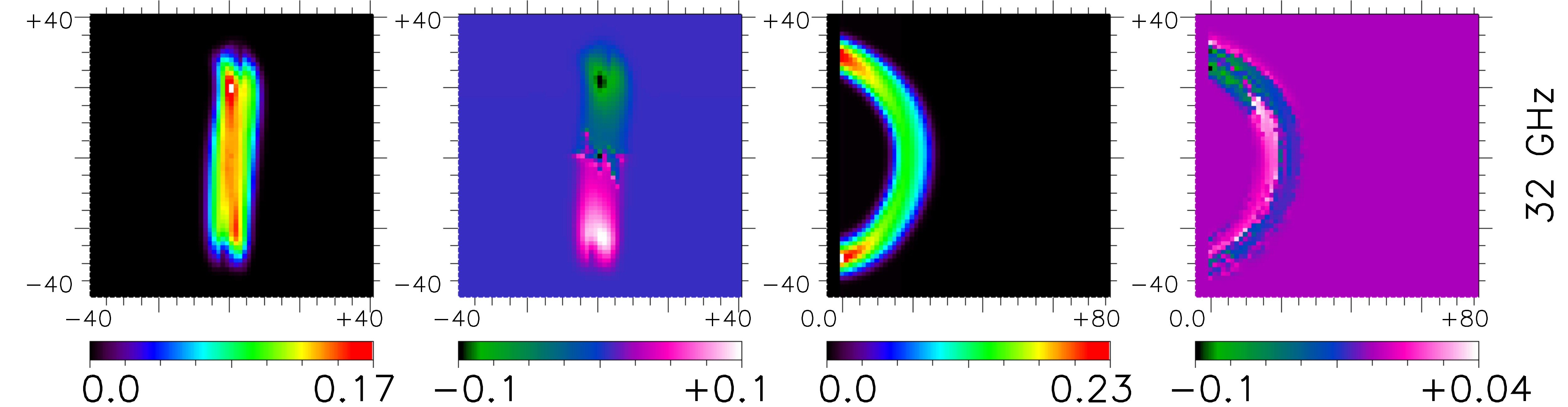}}
\caption{The same as in Figure~\ref{f-so}, but for the moment of reconnection decay (t=58~s).}
\label{f-sd}
\end{figure*}

\begin{table}
\centering
\caption{Parameters of thermal and non-thermal plasma taken from MHD and test-particle simulations of the reconnecting twisted loops for three different stages. Here, n$_b$ and $\gamma$ are the non-thermal electron density and power-law index of their energy spectrum. The lower and upper energy cut-offs are 10~keV and 1~MeV, respectively. Parameters n$_{L}$ and  T$_{L}$ are the thermal plasma density and temperature within the flaring loop, respectively, and R$_{fp}$ is the cross-section of the flaring loop near foot-points.}
\begin{tabular}{ l | l | l | l | l | l}
\hline 
Stage & n$_b$, 		  & $\gamma$ & R$_{fp}$, & n$_{L}$,  		& T$_{L}$,  \\
      &10$^{13}$m$^{-3}$  &          & Mm        & 10$^{15}$m$^{-3}$	& MK	    \\
\hline 
\multicolumn{6}{l}{Model A, Strong convergence, B$_{fp}$=1180~G} \\
Onset & 2.9 & 1.7 & 2.0 & 3.2 & 12 \\
Fast & 3.0 & 2.1 & 2.2 & 3.8 & 16 \\
Decay & 0.12 & 3.7 & 4.1 & 3.4 & 4 \\
\hline 
\multicolumn{6}{l}{Model B, Weak convergence, B$_{fp}$=320~G} \\
Onset & 4.1 & 1.5 & 3.2 & 3.6 & 18 \\
Fast & 2.7& 2.2 & 4.7 & 3.9 & 20 \\
Decay & 0.19 & 3.0 & 5.6 & 3.9 & 7 \\
\hline 
\multicolumn{6}{l}{Model C, Weak convergence, B$_{fp}$=680~G} \\
Onset & 8.2 & 1.6 & 3.2 & 3.3 & 30 \\
Fast & 4.2 & 2.0 & 4.8 & 3.8 & 36 \\
Decay & 0.64 & 3.5 & 5.6 & 3.9 & 15 \\
\hline 
\end{tabular}
\end{table}

\subsection{Calculation of synthetic microwave emission}

\begin{figure*}[ht!]    
\centerline{\includegraphics[width=0.7\textwidth,clip=]{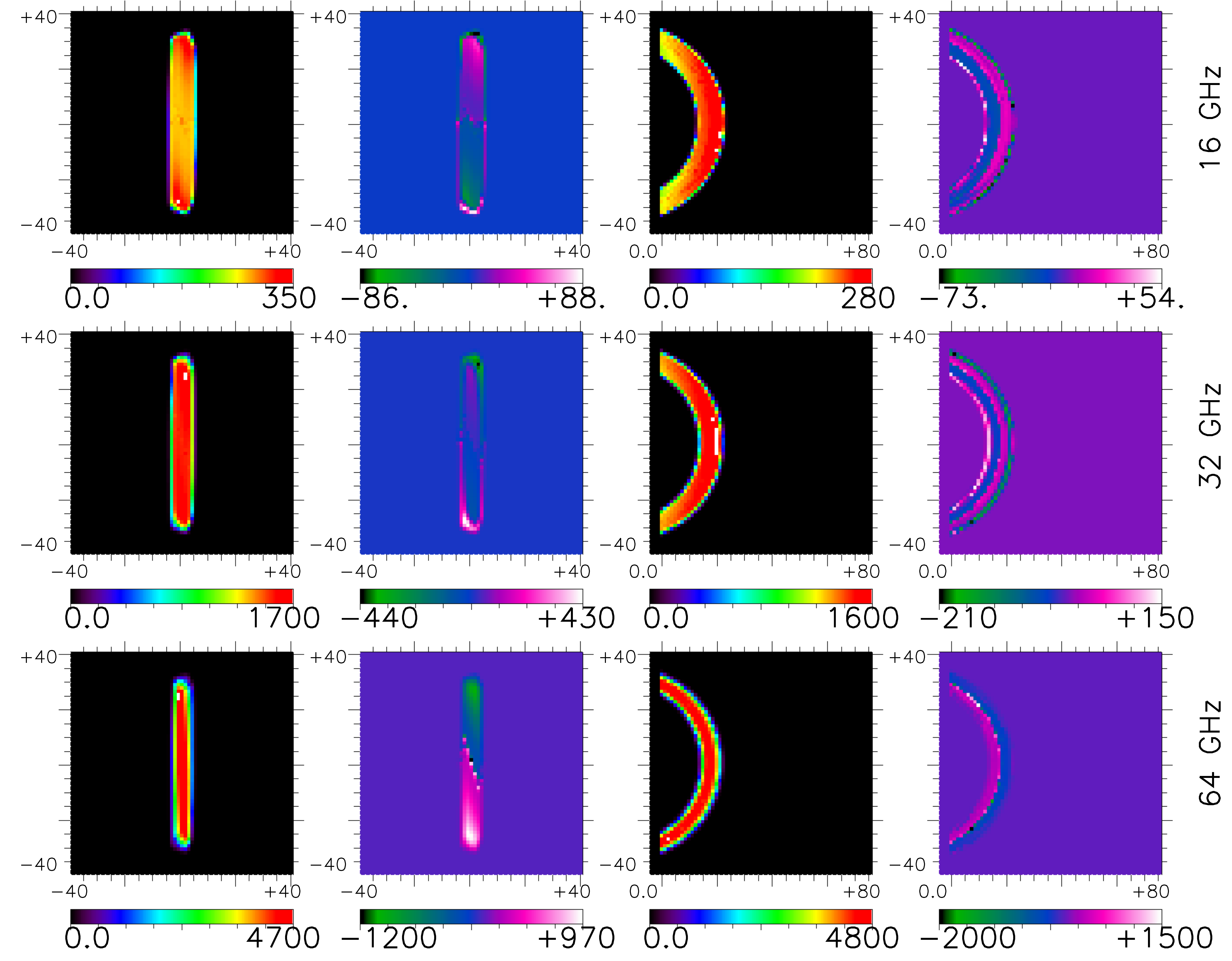}}
\caption{Microwave emission maps from the loop in model C soon after the kink instability and start of the magnetic energy release and particle acceleration (t=2~s). First and second columns correspond to Stokes I and V intensities, respectively, when the loop is seen from the top (X-Y plain). Third and fourth columns are Stokes I and V intensities, respecively, for the loop seen from its side (Z-Y plain). Different rows correspond to different frequencies. Intensities are given in sfu units.}
\label{f-vo}
\end{figure*}

Gyrosynchrotron microwave (GSMW) emission from our model twisted loops are calculated using the GX simulator \citep{nite15}, which is based on the fast aproximate calculation schemes developed by \citet{flku10}. The code can produce GSMW spectra and circular polarisation in the frequency range 1--95~GHz using the magnetic field and thermal plasma and non-thermal particle parameters. The magnetic field in each case is taken directly from MHD simulations. Plasma temperature and density outside the flaring loop are assumed to be constant with time. Their vertical distribution correspond to the initial conditions (undisturbed atmosphere) used in MHD simulations, the temperature in the chromosphere and the corona are about 33000~K and 4~MK, respectively, the desnity at the lower boundary (chromosphere) is $\sim 9 \times 10^{-8}$~kg~m$^{-3}$ ($5.6 \times 10^{13}$~cm$^{-3}$), and at 40~Mm (approximate loop top height) it is about $\sim 2 \times 10^{-12}$~kg~m$^{-3}$ (about $1.3 \times 10^{9}$~cm$^{-3}$). The low-temperature ambient plasma does not noticeably affect the emission maps, since plasma with temperatures below 1~MK does not emit or absorb microwave radiation at frequencies of interest. 

The flaring loop is formed by magnetic field lines originating from a circular foot-point, with the position and radius (R$_{fp}$, see Table 1) taken in each case from MHD simulations. The temperature and density of plasma in the flaring loop are assumed to be nearly constant (they change to the ambient values in a thin layer close to the loop surface), their values (Table 1) approximately correspond to those in MHD simulations. The non-thermal electron population is nearly uniform within the loop (its density quickly drops to zero at the loop surface) and has spatially uniform energy and pitch-angle distribution. Their energy spectrum is a single power-law, while the pitch-angle distribution corresponds to one of the three cases (see Figure~\ref{f-pangle}).  The non-thermal electron parameters are approximated from the test-particle simulations, assumed to be as given in the Table 1 for three different models. The model A has strongly converging field (convergence factor 10, and, hence, loop-top to foot-point cross-section ratio of about 3.2), while models B and C have weakly converging loops (convergence factor 2 and loop-top to foot-point cross-section ratio of 1.4). The average (along the loop) magnetic field in model A is similar to that in model C. 

We consider three different instants in time for each model: one just after the kink-instability  (i.e. just after the magnetic reconnection, energy release and particle acceleration begin, t=2~s), another corresponding to the middle of the relaxation process (approximately, during the temperature peak, t=30~s), and, finally, one corresponding to the decay of energy release (t=58~s). The whole relaxation phase (about 60~s long in our models) represents the impulsive phase in solar flares. Because of the uncertainty with pitch-angle distribution, most of the maps Figures~\ref{f-so}-\ref{f-vth}) are calculated for isotropic distributions, and later compared with collimated and pancake-like distributions (Figure~\ref{f-paeffect}).

\section{Results and discussion}\label{mwave}

Volume integrated GSMW spectra for one of the models (A) are shown in Figure~\ref{f-gsmwspectra}. The total intensities are quite higha and would correspond to a major flare; this is due to considerably hard electron energy spectra. Furthermore, because of the hard electron spectra, the maxima in GSMW spectra appear at relatively high frequencies; thus, during the fast energy release the maximum drifts from about 15-20~GHz to 8-10~GHz.

\begin{figure*}[ht!]    
\centerline{\includegraphics[width=0.7\textwidth,clip=]{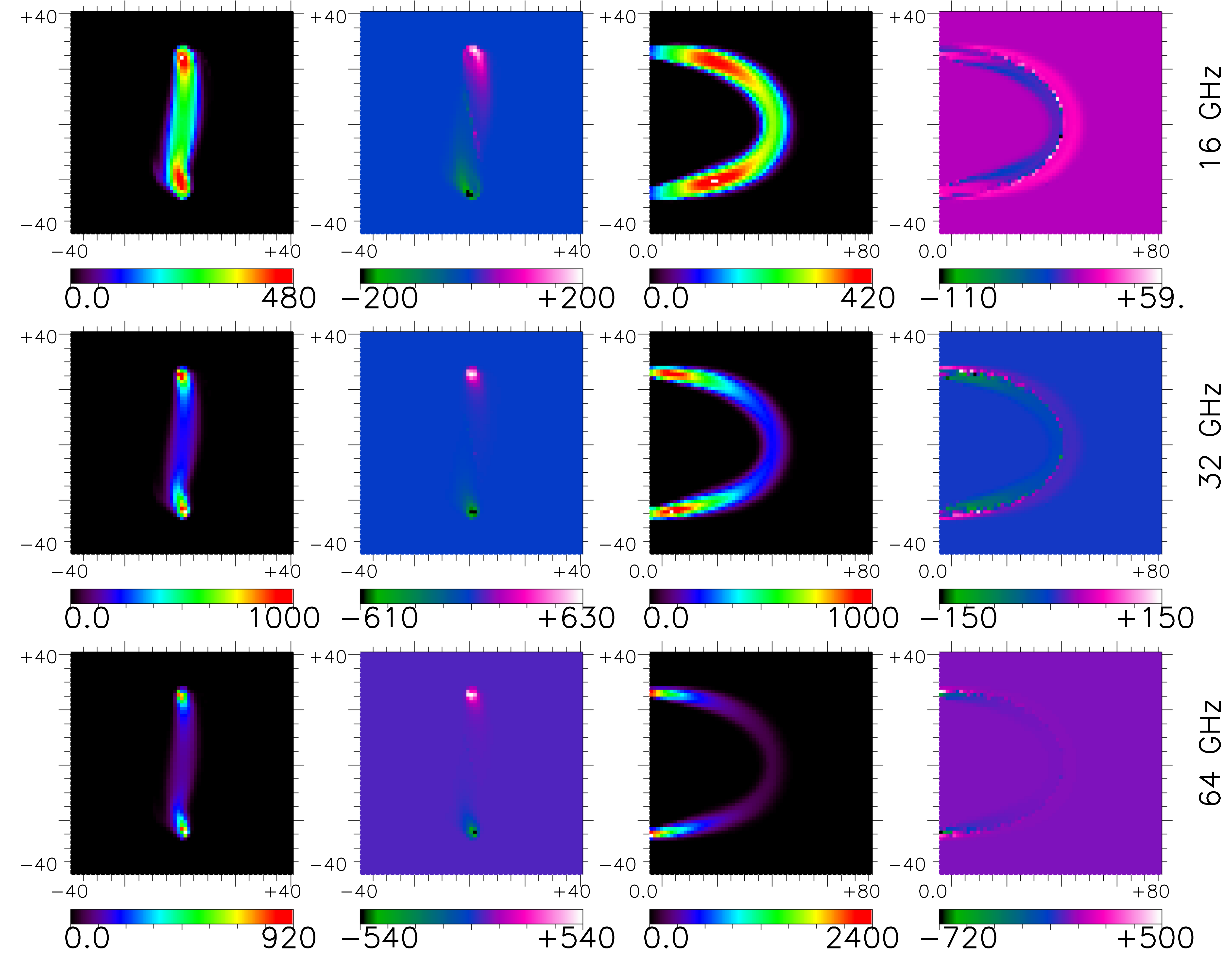}}
\caption{Microwave emission maps from the loop in model A soon after the kink instability and start of the magnetic energy release and particle acceleration (t=2~s). First and second columns correspond to Stokes I and V intensities, respectively, when the loop is seen from the top (X-Y plain). Third and fourth columns are Stokes I and V intensities, respecively, for the loop seen from its side (Z-Y plain). Different rows correspond to different frequencies. Intensities are given in sfu units.}
\label{f-zo}
\end{figure*}

\begin{figure*}[ht!]    
\centerline{\includegraphics[width=0.7\textwidth,clip=]{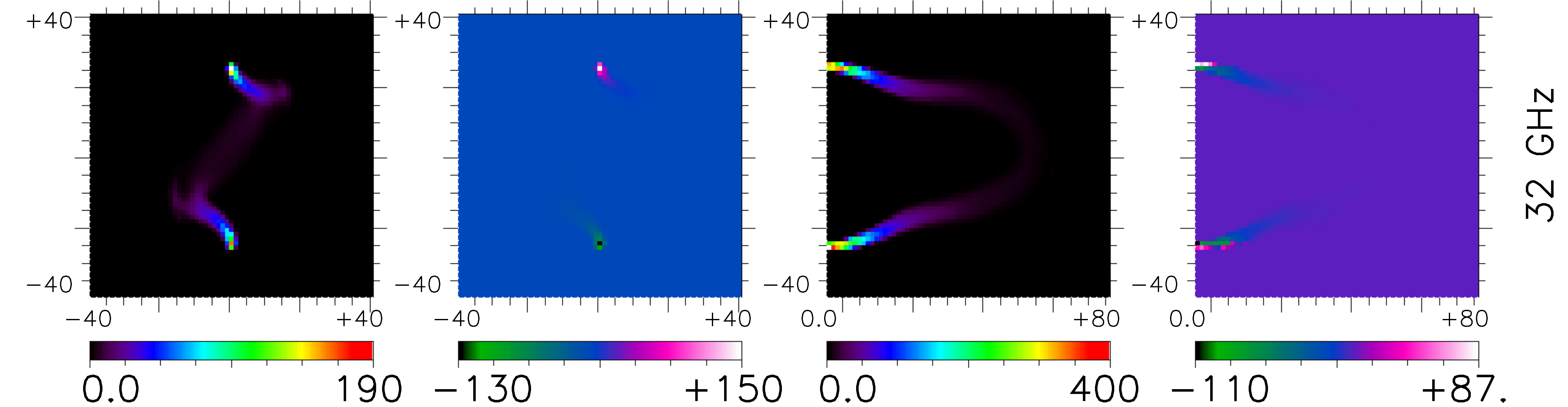}}
\caption{The same as in Figure~\ref{f-zo} but for t=30~s.}
\label{f-zf}
\end{figure*}

\begin{figure*}[ht!]    
\centerline{\includegraphics[width=0.7\textwidth,clip=]{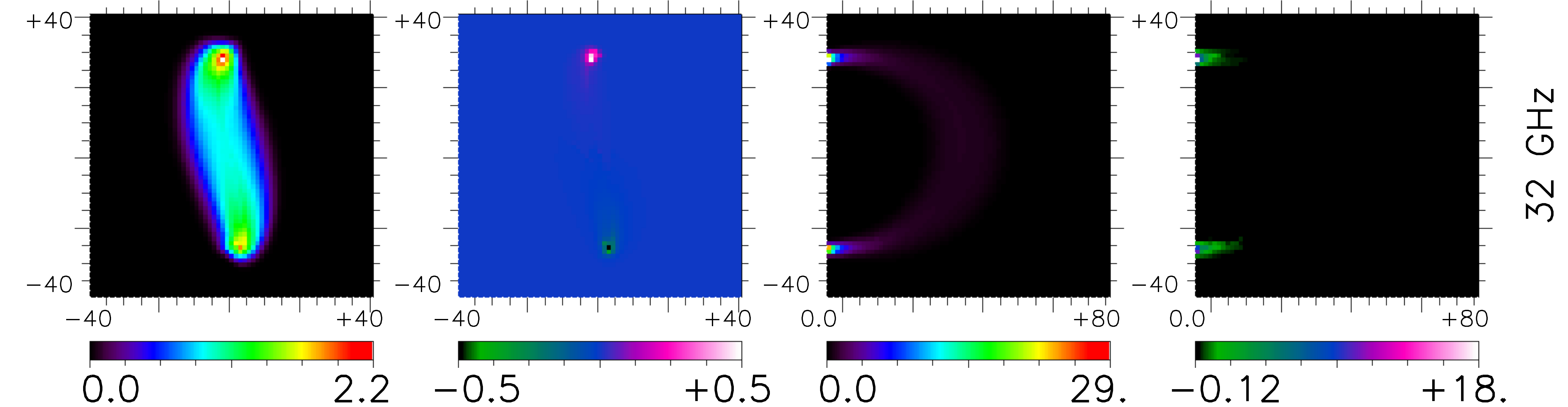}}
\caption{The same as in Figure~\ref{f-zo} but for t=58~s. {\bf Note, that intensities are given in 10$^{-3}$~sfu unites.}}
\label{f-zd}
\end{figure*}

\begin{figure}[ht!]    
\centerline{\includegraphics[width=0.35\textwidth,clip=]{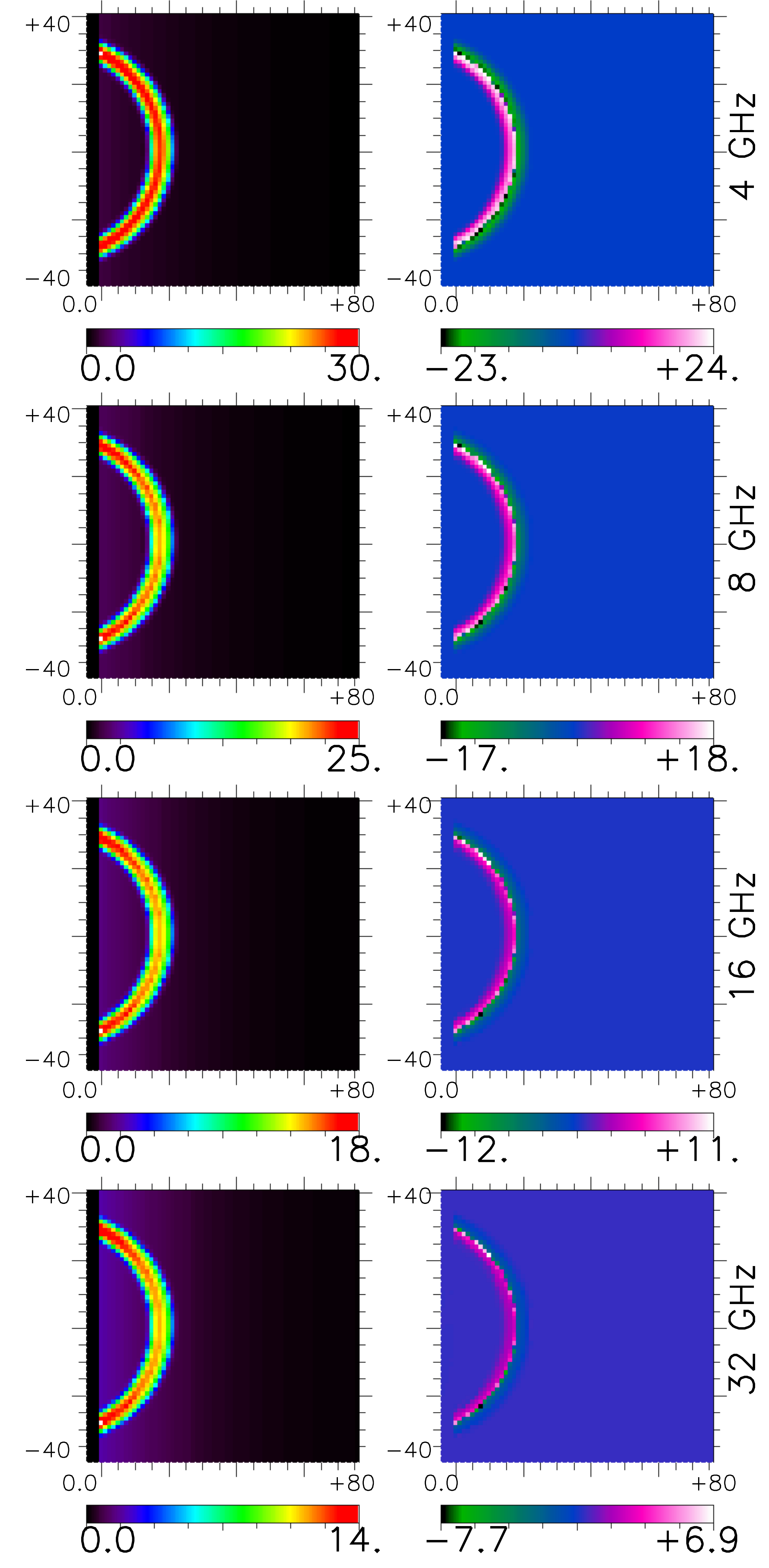}}
\caption{Microwave emission produced only by thermal plasma in the loop in model C soon after the kink instability  (t=2~s). The loop is seen from its side (Z-Y plain). First and second columns correspond to Stokes I and V intensities, respectively. Different rows correspond to different frequencies shown on the right. {\bf Note, that intensities here are given in 10$^{-3}$~sfu units.} }
\label{f-vth}
\end{figure}

The synthetic microwave emission and polarisation maps from twisted loops containing high-energy electrons are shown in Figures~\ref{f-so}-\ref{f-vth}.  First, we discuss the frequency variation of the microwave intensity and polarisation in our simulations. The emission is optically thick at lower frequencies (below 2-20~GHz, depending on the non-thermal electron spectral index and magnetic field value), and optically thin at higher frequencies. Typically, the intensities increase from few sfu at 4~GHz to $\sim$10$^3$ above 64~GHz just after reconnection begins, while the polarization $V/I$ increases from 25-50\% at lower frequencies to 10-20\% at higher frequencies. It needs to be noted, that some of our intensities are quite high, corresponding to major flares. This is likely to be due to hard electron spectra resulting from our test-particle simulations. Indeed, a particle population with the spectral index of 3 and the lower energy cut-off 10~keV would have approximately 100 times less energetic electrons at 1~MeV, compared to the current electron ppopulation with the spectral index of 
$\sim$~2. In reality, we would expect the absolute intensities of microwave emission from smaller flares in confined loops to be substantially weaker.

The hardness of the particle spectra affects not only total microwave intensities, but also the shapes of the spectra. Most importantly, it shifts the spectral peaks, effectively dividing optically-thck and thin spectral ranges towards higher frequencies. At the same time, it should be noted, that our spectra are not unnatural, there are plenty of flares with the spectral index of 2-3 \cite[e.g.][]{kawe12}, i.e. comparable to the simulated particle models in Sect.2.

The intensities decrease with time, as the energetic electron numbers decrease. However, because of softening electron energy spectra, the GSMW spectra also change, and the intensity decreases faster at higher frequencies. Thus, in model A, the intensity at 32~GHz decreases from about 1900~sfu at the beginning of relaxation, to about 90~sfu around the middle of the relaxation and then drops to 0.2~sfu towards the end of energy release. At the same, the intensities at 16~GHz drop from about 500~GHz to 120~GHz during the first half of the relaxation, and then also drop to $<$1~sfu towards the end of the energy release. The polarisation degrees remain nearly the same during evolution of these loops.

The most interesting question, of course, is the spatial distribution of GSMW polarisation. It is sensitive to the frequency, mostly because the loops are optically thick at lower frequencies, but thin at high frequencies. In the weakly-converging loops (Models B and C), the CLPG patterns can be clearly seen at higher frequencies, above 30~GHz in model B and above 60~GHz in model C (the difference is due to higher magnetic field in model C). Thus, they are visible near the looptops when loops are observed from the top, and along whole loops, when loops are observed from the side. At lower frequencies it is more complicated. When these loops are observed from the top, only emission coming from the edges of microwave sources is optically thin and polarised according to the sign of LOS magnetic field. However, the emission produced at lower frequencies within the source is optically thick and is polarised oppositely to the optically thin emission, due to the self-absorption. It is practically impossible to see the CLPG pattern at lower frequencies in loops observed from the top. Loops observed from the side at low frequencies show a polarisation gradient, however, it more more complicated, with several sign reversals.

The cross-sections of the considered loops are about 8-12~Mm near loop tops. Therefore, in order to detect CLPG patterns in such loops, the angular resolution needs to be about 10~arcsec ol better.

Now, let us consider the effect of loop geometry. It appears that CLPG patterns would be more difficult to see in strongly converging loops. This is because their high magnetic field variation from loop-tops to foot-points means that most of GSMW emission comes from foot-points, which are by more than one order-of-magnitude brighter than the loop-tops at higher frequencies. Thus, in model A (Figure~\ref{f-zo}, see also Figures \ref{f-zf}-\ref{f-zd} in Appendix) the intensity drops from about 2.4$\times$10$^3$~sfu near the footpoints to about 200~sfu around loop-top at 64~GHz, and from about 1000~sfu to 120~sfu at 32~GHz. As the result, at higher frequencies GSMW emission from a strongly converging loop observed from the top looks like two foot-point sources, similar to the HXR emission. The two foot-points have opposite circular polarisation, as expected, with no visible structure. At lower frequencies, however, the emission is more extended, and CLPG pattern can be seen from the top, although the intensity near the polarisation reversal is low. When a strongly converging loop is seen from the side, the CLPG pattern can be seen on a wider frequency range. Thus, in model A, CLPG can be seen along the whole loop at 16~GHz (although, the polarisation structure is complicated near the optically thick foot-points), as well as at 32 and 64~GHz, although at higher frequencies that intensity of the loop top is very low. 


It can be seen that in all considered cases, these CLPG patterns are best visible just after the onset of magnetic reconnection (see Figure~\ref{f-so}, \ref{f-zo} and \ref{f-vo}). In weakly-converging loops B and C observed from the top, the polarisation reversal line is nearly perpendicular to the loops towards the middle of the energy release (when the temperature peaks). This, obviously, is a result of twist reduction. However, the CLPG patterns still can be seen when the loops are observed from the side, although they are fading and the patterns become more complicated due to the optical thickness effects and because the loops lose their rope-like structure. Hence, in the weakly-converging loops, the CLPG structure can be observed for about half of the impulsive phase (around 30~s in our simulations) and for up to 60~s in loops observed from the side. Strongly-converging loops lose their rope-like structure faster and their field quickly becomes more chaotic. As the result, during the later stages of energy release in loop A (Figures~\ref{f-zf}-\ref{f-zd}) the CLPG pattern cannot be seen. Therefore, in strongly-converging loops the CLPG pattern can be observed for less than a half of impulsive phase, or about 10-20~s in our simulations.

Of course, the lifetime of the CLPG patterns strongly depends on the time-scale of the loop evolution. Generally, it might be possible to have twisted loops which are not evolving, for instance, when a twisted loop is a part of a complex Active Region, where magnetic reconnection, energy release and particle acceleration occur outside that loop. In this case, if energetic particles manage to get into the loop, the twist would result in the CLPG polarisation pattern; however, because the loop is not evolving and the twist does not dissapear, the lifetime of the CLPG pattern would not be limited. Another possible scenario is when the microwaves are produced by hot thermal plasma in non-evolving or very slowly-evolving loop (for instance, due to slow current dissipation or localised magnetic reconnection near a foot-point without twist reduction). {There are several observations of microwave emission produced by thermal electrons in hot flaring plasma \cite[e.g.][]{gahu89, flee15}.} In this case, again, the lifetime of the CLPG pattern could be much longer. Indeed, thermal GSMW radiation produced by plasma with temperatures of 10-20~MK clearly demonstrates the CLPG patterns (Figure~\ref{f-vth}). The polarisation degrees can be substantially higher that in non-thermal GSMW in our models: $V/I$ increases from 10\% at 95~GHz to 80\% at 4~GHz. However, the intensities are very low, from about 0.02 at 4~GHz down to 0.003~sfu at 90~GHz.

\begin{figure}[ht!]
\centerline{\includegraphics[width=0.35\textwidth,clip=]{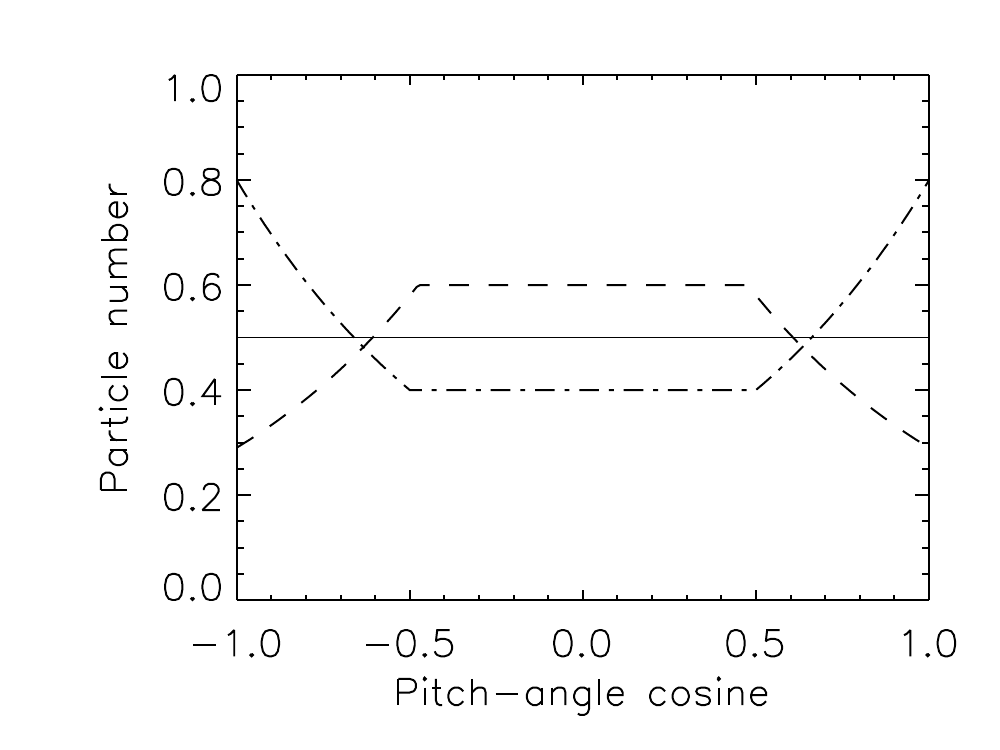}}
\caption{Pitch-angle distribution of energetic electrons in the loop: solid -- isotropic, dashed -- collimated distribution, dot-dashed -- exponential loss-cone.}
\label{f-pangle}
\end{figure}

\begin{figure}[ht!]
\centerline{\includegraphics[width=0.35\textwidth,clip=]{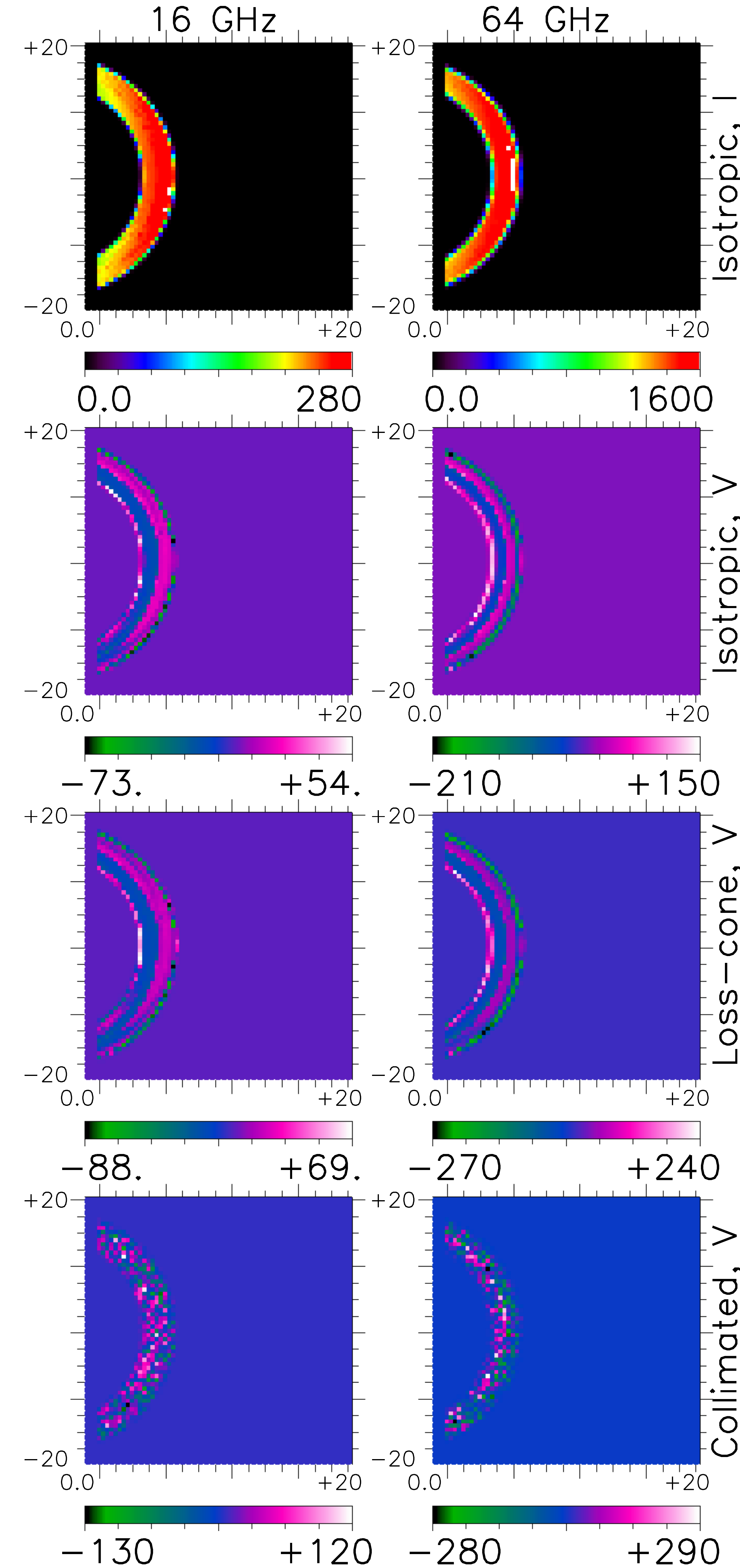}}
\caption{Effect of pitch-angle distribution of the CLPG pattern visibilities in Model C (t=2~s). Different rows correspond to different models, different columns correspond to different frequencies.}
\label{f-paeffect}
\end{figure}

Finally, the pitch-angle distribution of energetic electrons is a very important issue \cite[see e.g.][]{rama69,flme03, sico10,kuze11}. Therefore, we calculated emission and polarisation maps for different types of pitch-angle distributions (Figure~\ref{f-pangle}): collimated, $\frac{N(\cos\theta = \pm 1)}{N(\cos\theta = 0)} = 2$, and exponential loss-cone, $\frac{N(\cos\theta = \pm 1)}{N(\cos\theta =  0)} = 0.5$, and compare them with those produced by isotropic distribution. Corresponding polarisation maps are compared in Figure~\ref{f-paeffect}. It can be seen that the loss-cone distributions produce polarization patterns similar to those produced by isotropic distribution (described above), both qualitatively and quantitatively. The emission is optically thick at lower frequencies (below $\sim$40~GHz) and optically thin at higher frequencies. The CLPG pattern is seen at different frequencies, although it is opposite when the emission is optically thick. In contrast, the collimated distribution appears to produce optically thin emission, however, both intensity and polarisation maps are much noisier. The CLPG gradient can be seen at lower frequencies (below 30-35~GHz) in the loops observed from the side, but completely sinks in the noise at higher frequencies and when the loops are observed from the top. This effect is not unexpected: indeed, electrons with pitch-angles $\approx \pm 1$ produce very little GSMW emission; \mgc{however the pitch-angles may become high in regions with high magnetic field curvature, producing localised emissivity spikes and, hence, producing very noisy emission maps}.

\section{Summary}\label{summ} 

Our results show that, indeed, twisted coronal loops can produce characteristic microwave polarisation (CLPG) patterns -- a noticeable Stokes V component gradient across the loop, confirming earlier studies by \citet{shku16}. However, this pattern would be visible only in some cases, depending on the magnetic field, thermal and non-thermal plasma parameters, loop orientation and magnetic field geometry.

The CLPG pattern is more clear in loops seen from the side ({\it e.g.} when observed very close to the limb). Thus, in weakly converging twisted loops (models B and C just after the instability) observed from side the pattern can be seen along the whole body of the loop, at least at higher frequencies ($>$30-60~GHz, depending on the magnetic field strength) where GSMW emission is optically thin. When these loops are observed from the top, the pattern can be seen, but it would require higher spatial resolution to be detected {compared to the observation from loop's side}. In strongly converging loops (model A), the CLPG pattern can be seen when the loop is observed from the side. When the loop is observed from the top, it can be seen only at lower frequencies (below 15-20~GHz), because at higher frequencies the loop-top emission is very weak.

The pitch-angle distribution of energetic particles strongly affects the visibility of the pattern: CLPG is clearly seen in microwave produced by an isotropic or loss-cone (pancake-like) electron distributions, while being very noisy when produced by electrons collimated along magnetic field. Still, the CLPG pattern can be seen, at least at lower frequencies (below 15-20~GHz).

The CLPG pattern in microwaves produced by flaring loops is a transient phenomenon, its lifetime is shorter than the length of the impulsive phase. Thus, in weakly-converging loops its duration is about half of the impulsive phase duration (about 30~s in our simulations). Interestingly, the pattern can be seen for a longer time in loops observed from the side. In the strongly-converging loops, the lifetime of the CLPG patterns is even shorter, about a third of the impulsive phase. In our models, these patterns disappear within 10-20~s after kink instability. 

In principle, the CLPG patterns can live much longer if produced by (relatively) non-evolving loops. This could be the case when either energetic particles or hot plasma are injected into a twisted loop from outside, for instance due to magnetic reconnection near the foot-points (without loss of twist). Thermal GSMW appears to produce very clear CLPG patterns, although the microwave intensity is very low, about 10$^{-2}$~sfu, compared to about 10$^2$~sfu, produced by non-thermal electrons.

We conclude that CLPG patterns can be seen with spatial resolution of about 10~arcsec or better (depending on the loop size) and, hence, can be detected using instruments such as Nobeyama radioheliograph (at 17~GHz, where polarimetry is possible) and future solar SKA.

\begin{acknowledgements}
\mrk{We are thankful to Gregory Fleishman (NJIT) and Alexei Kuznetsov (ISTP Irkutsk) for their help with GX simulator. This work is funded by Science and Technology Facilities Council (STFC) (grants ST/L000768/1, ST/L000741/1). MHD simulations have been performed using STFC funded DiRAC Data Centric system at Durham University (grants ST/K00042X/1, ST/K00087X/1, ST/K003267/1).}
\end{acknowledgements}


\begin{thebibliography}{00}
\bibitem[\protect\citeauthoryear{Alexander et al.}{2006}]{alee06} Alexander, D., Liu, R. and Gilbert, H.R., 2006, ApJ, 653, 719.
\bibitem[\protect\citeauthoryear{Arber et al.}{2001}]{arbe01} Arber, T.G., Longbottom, A. W., Gerrard, C. L. \& Milne, A. M. 2001, J.Comp.Phys, 171, 151
\bibitem[\protect\citeauthoryear{Aschwanden et al.}{2009}]{asce09} Aschwanden, M.J., Wuelser, J.P., Nitta, N.V. \& Lemen, J.R. 2009, Solar Phys., 256, 3
\bibitem[\protect\citeauthoryear{Bareford et al.}{2016}]{bare16} Bareford, M. R., Gordovskyy, M., Browning, P.K. \& Hood, A.W. 2016, Solar Phys, 291, 187.
\bibitem[\protect\citeauthoryear{Cargill et al.}{2012}]{care12} Cargill, P.J., Vlahos, L., Baumann, G., Drake, J.F. \& Nordlund, A., 2012, Space Sci. Rev., 173, 223.
\bibitem[\protect\citeauthoryear{Doschek et al.}{2007}]{dose07} Doschek, G.A., Mariska, J.T., Warren, H.P., Brown, C.M., Culhane, J.L., Hara, H., Watanabe, T., Young, P.R. \& Mason, H.E., 2007, ApJ Lett., 667, L109.
\bibitem[\protect\citeauthoryear{Doschek et al.}{2008}]{dose08} Doschek, G.A., Warren, H.P., Mariska, J.T., Muglach, K., Culhane, J.L., Hara, H. \& Watanabe, T., 2008, ApJ, 686, 1362
\bibitem[\protect\citeauthoryear{Fan}{2010}]{fan10} Fan, Y., 2010, ApJ, 719, 728.
\bibitem[\protect\citeauthoryear{Fleishman \& Kuznetsov}{2010}]{flku10} Fleishman, G.D. and Kuznetsov A.A., 2010, ApJ, 721, 1127.
\bibitem[\protect\citeauthoryear{Fleishman \& Melnikov}{2003}]{flme03} Fleishman, G.D. and Melnikov, 2003, ApJ, 587, 823.
\bibitem[\protect\citeauthoryear{Fleishman et al.}{2015}]{flee15} Fleishman, G.D., Nita, G.M. \& Gary, D.E., 2015, ApJ, 802, 122.
\bibitem[\protect\citeauthoryear{Gary \& Hurford}{1989}]{gahu89} Gary, D.E. and Hurford, G.J., 1989, ApJ, 339, 1115.
\bibitem[\protect\citeauthoryear{Gary et al.}{2013}]{gare13} Gary, D.E., Fleishman, G.D. \& Nita, G.M., 2013, Solar Phys., 288, 549.
\bibitem[\protect\citeauthoryear{Gimenez de Castro et al.}{2009}]{gime09} Gimenez de Castro, C.G., Trottet, G., Silva-Valio, A., Krucker, S., Costa, J.E.R., Kaufmann, P., Correia, E., Levato, H., 2009, A\&A, 507, 433.
\bibitem[\protect\citeauthoryear{Gordovskyy \& Browning}{2011}]{gobr11} Gordovskyy, M. \& Browning, P.K. 2011, ApJ, 729, 101
\bibitem[\protect\citeauthoryear{Gordovskyy \& Browning}{2012}]{gobr12} Gordovskyy, M. \& Browning, P.K. 2012, Solar Phys., 277, 299
\bibitem[\protect\citeauthoryear{Gordovskyy et al.}{2014}]{gore14} Gordovskyy, M., Browning, P.K., Kontar, E.P. \& Bian, N.H. 2014, A\&A, 561, 72.
\bibitem[\protect\citeauthoryear{Gordovskyy et al.}{2016}]{gore16} Gordovskyy, M., Kontar, E.P. \& Browning, P.K. 2016, A\&A, 589, 104.
\bibitem[\protect\citeauthoryear{Gordovskyy et al.}{2010}]{gore10} Gordovskyy, M., Browning, P.K. \& Vekstein G.E. 2010, ApJ, 720, 1603
\bibitem[\protect\citeauthoryear{Hanaoka}{2005}]{hana05} Hanaoka, Y., 2005, PASJ, 57, 235.
\bibitem[\protect\citeauthoryear{Iwai \& Shibasaki}{2013}]{iwsh13} Iwai, K. and Shibasaki, K., 2013, PASJ, 65, S14.
\bibitem[\protect\citeauthoryear{Jeffrey \& Kontar}{2013}]{jeko13} Jeffrey, N.L.S. and Kontar E.P., 2013, ApJ, 766, 75.
\bibitem[\protect\citeauthoryear{Karlicky \& Kliem}{2010}]{kakl10} Karlicky, M. and Kliem, B., 2010, Solar Phys, 266, 71.
\bibitem[\protect\citeauthoryear{Kawate et al.}{2012}]{kawe12} Kawate, T., Nishizuka, N., Oi, A., Ohyama, M. \& Nakajima, H., 2012, ApJ, 747, 131.
\bibitem[\protect\citeauthoryear{Kontar et al.}{2011}]{kone11} Kontar, E.P., Hannah, I.G. and Bian, N.H., 2011, ApJ Lett., 730, L22.
\bibitem[\protect\citeauthoryear{Kucera et al.}{1993}]{kuce93} Kucera, T. A., Dulk, G.A., Kiplinger, A.L., Winglee, R.M., Bastian, T.S.\& Graeter, M., 1993, Solar Phys., 412, 853.
\bibitem[\protect\citeauthoryear{Kuridze et al.}{2013}]{kure13} Kuridze, D., Mathioudakis, M., Kowalski, A.F., Keys, P.H., Jess, D.B., Balasubramaniam, K.S. and Keenan, F.P., 2013, A\&A, 552, A55.
\bibitem[\protect\citeauthoryear{Kuznetsov et al.}{2011}]{kuze11} Kuznetsov, A.A., Nita, G.M. \& Fleishman, G.D., 2011, ApJ, 742, 87.
\bibitem[\protect\citeauthoryear{Nindos et al.}{2000}]{nine00} Nindos, A., White, S.M., Kundu, M.R. \& Gary, D.E., 2000, ApJ, 533, 1053.
\bibitem[\protect\citeauthoryear{Nita et al.}{2015}]{nite15} Nita, G.M., Fleishman, G.D., Kuznetsov, A.A., Kontar, E.P. and Gary, D.E., 2015, ApJ, 799, 236.
\bibitem[\protect\citeauthoryear{Petrosian}{1981}]{petr81} Petrosian, V., 1981, ApJ, 251, 727.
\bibitem[\protect\citeauthoryear{Pinto et al.}{2016}]{pine16} Pinto, R., Gordovskyy, M., Browning, P.K. \& Vilmer, N. 2016, A\&A, 585, 159.
\bibitem[\protect\citeauthoryear{Ramaty}{1969}]{rama69} Ramaty, R., 1969, ApJ, 158, 753.
\bibitem[\protect\citeauthoryear{Reznikova et al.}{2015}]{reze15} Reznikova, V.E., Van Doorsselaere, T. \& Kuznetsov, A.A., 2015, A\&A, 575, A47.
\bibitem[\protect\citeauthoryear{Sharykin \& Kuznetsov}{2016}]{shku16} Sharykin, I.N. and Kuznetsov, A.A., 2016, Solar Phys., 291, 1341.
\bibitem[\protect\citeauthoryear{Simoes \& Costa}{2010}]{sico10} Simoes, P.J.A. \& Costa, J.E.R., 2010, Solar Phys., 266, 109.
\bibitem[\protect\citeauthoryear{Shibata et al.}{1995}]{shie95} Shibata, K., Masuda, S., Shimojo, M., Hara, H., Yokoyama, T., Tsuneta, S., Kosugi, T. and Ogawara, Y., 1995, ApJ Letters, 451, L83.
\bibitem[\protect\citeauthoryear{Srivastava et al.}{2010}]{srie10} Srivastava, A.K., Zaqarashvili, T.V., Kumar, P. \& Khodachenko, M.L. 2010, ApJ, 715, 292
\bibitem[\protect\citeauthoryear{Tzatzakis et al.}{2008}]{tzae08} Tzatzakis, V., Nindos, A. \& Alissandrakis, C.E., 2008,  Solar Phys., 253, 79.
\end{thebibliography}
\end{document}